\documentclass[12pt]{iopart}

\usepackage{iopams}  
\usepackage{setstack}

\usepackage{subfigure}
\usepackage{subeqn}
\usepackage{graphicx}

\graphicspath{{./}{./images/}{./images_all/}{./images_all/Landauer_limit/}}
\DeclareGraphicsExtensions{.eps}

\begin{document}

\title[]{Landauer limit of energy dissipation in a magnetostrictive particle}

\author{Kuntal Roy}
\address{School of Electrical and Computer Engineering, Purdue University, West Lafayette, Indiana 47907, USA\\
	* Some works for this paper were performed prior to joining Purdue University.}

\ead{royk@purdue.edu}

\date{\today}

\begin{abstract}
According to Landauer's principle, a minimum amount of energy proportional to temperature must be dissipated during the erasure of a classical bit of information compensating the entropy loss, thereby linking the information and thermodynamics. Here we show that the Landauer limit of energy dissipation is achievable in a shape-anisotropic single-domain magnetostrictive nanomagnet having two mutually anti-parallel degenerate magnetization states that store a bit of information. We model the magnetization dynamics using stochastic Landau-Lifshitz-Gilbert equation in the presence of thermal fluctuations and show that on average the Landauer bound is satisfied, i.e., it accords to the generalized Landauer's principle for small systems with stochastic fluctuations. 

\end{abstract}


\maketitle

\section{Introduction} Computers are physical systems and a computation or processing of information whether performed in an electronic machinery or in a biological system is subjected to physical principle, e.g., the second law of thermodynamics. According to Landauer's principle~\cite{RefWorks:148}, in a classical system with two degenerate ground states, it necessarily dissipates an energy of amount $kT ln(2)$ ($\sim3 \times 10^{-21}$ joule at T=300 K, $k$ is the Boltzmann constant, and T is temperature) to erase a bit of information (logically \emph{irreversible} computation contrary to reversible computation~\cite{RefWorks:523,RefWorks:436,RefWorks:625,RefWorks:654,RefWorks:653,RefWorks:646}), which compensates for the entropy loss. In fact, the intimate relationship between information and entropy dates back over a century ago when Maxwell introduced his famous and controversial \emph{demon}~\cite{maxwell,RefWorks:615,RefWorks:619,RefWorks:642}, which was later exorcized by Bennett~\cite{RefWorks:614,RefWorks:627,RefWorks:528,RefWorks:527,RefWorks:645,RefWorks:613,RefWorks:522,RefWorks:144,RefWorks:644}. However, the possibility of stochastic violation of Landauer's principle cannot be discarded in small systems that are prone to thermal fluctuations~\cite{RefWorks:629,RefWorks:637,RefWorks:626,RefWorks:623,RefWorks:624,RefWorks:643}. Indeed the stochastic violations of second law have been experimentally observed~\cite{RefWorks:621,RefWorks:622}, however, \emph{on average} the second law is still safeguarded~\cite{RefWorks:620}. There are controversies whether Landauer's bound holds for asymmetric memories~\cite{RefWorks:616,RefWorks:658,RefWorks:660} and in quantum regime~\cite{RefWorks:656,RefWorks:657}, although the second law of thermodynamics remains intact.

Recently, Landauer limit of energy dissipation is experimentally demonstrated for a colloidal particle with \emph{linear} motion trapped in a double-well potential landscape utilizing a laser beam~\cite{RefWorks:617}. The heat dissipated during the erasure process (i.e., while resetting the particle in one of the two wells) is measured and shown to be approaching the Landauer's limit. However, the Landauer's limit is not experimented for a \emph{rotational} body like the magnetization in a nanomagnet~\cite{RefWorks:557,RefWorks:157,RefWorks:490,RefWorks:133,RefWorks:402,RefWorks:426}. We consider the complete three-dimensional potential landscape and full three-dimensional motion of the magnetization, unlike \emph{assuming} an overdamped particle with \emph{linear} motion~\cite{RefWorks:623,RefWorks:624} and show that the magnetization may deflect out of magnet's plane during its dynamical motion. Consideration of even a small out-of-plane excursion of magnetization plays a crucial role in shaping the magnetization dynamics while achieving the Landauer limit of energy dissipation. Here, we particularly consider a single-domain \emph{magnetostrictive} nanomagnet and show that Landauer limit of energy dissipation is achievable in such a magnetostrictive particle. Considering that the excessive energy dissipation during switching of bits has been the bottleneck behind utilizing traditional charge-based transistor electronics further~\cite{kilby,moore65,RefWorks:211}, magnetostrictive nanomagnets in multiferroic heterostructures have potential to become the staple of modern non-volatile memory and logic systems~\cite{RefWorks:164,roy11,roy11_2,roy13_2,roy11_6,roy13_spin,roy13,roy14,RefWorks:649,roy14_2,RefWorks:559,RefWorks:609,RefWorks:611,RefWorks:790}.  We have solved stochastic Landau-Lifshitz-Gilbert (LLG) equation~\cite{RefWorks:162,RefWorks:161,RefWorks:186} in the presence of thermal fluctuations to theoretically analyze the aspects of magnetization dynamics and determine the energy dissipation during the erasure of a bit of information. Simulation results also show that stochastic violation of the Landauer bound is possible, nonetheless the bound is respected on average.

\begin{figure*}[htbp]
\centering
\includegraphics[width=170mm]{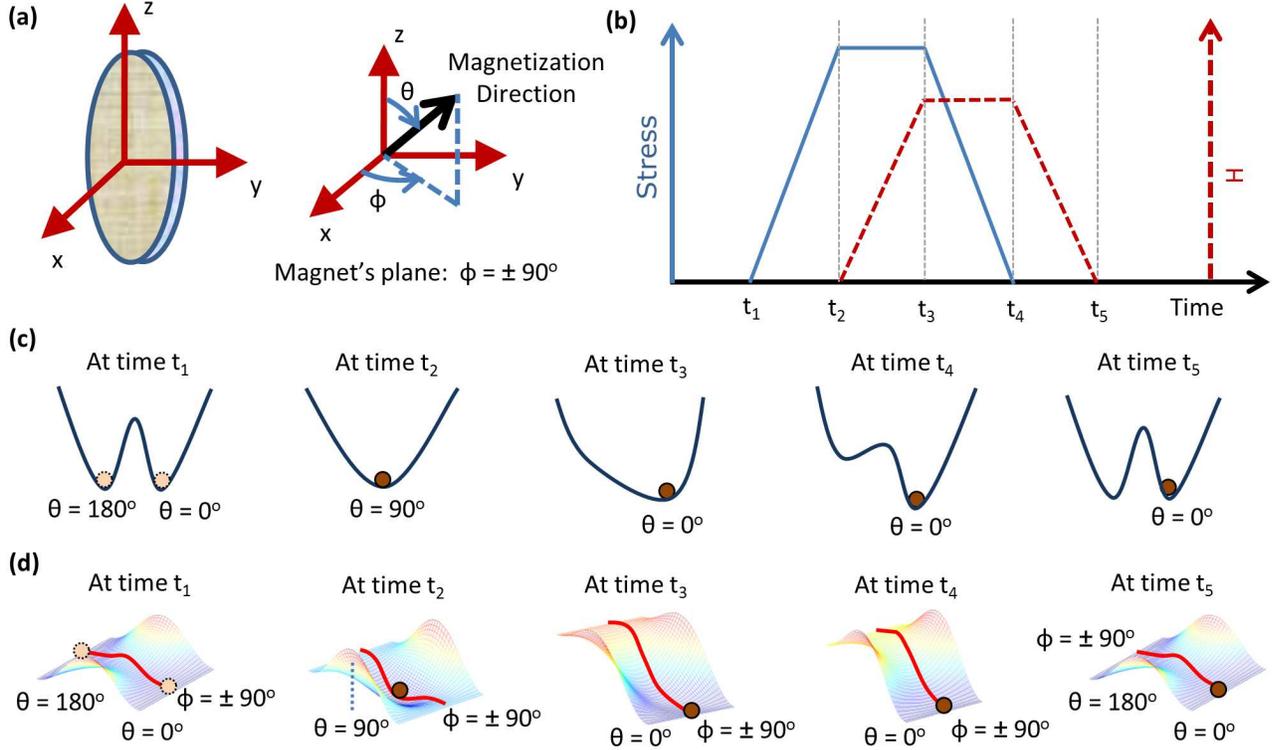}
\caption{(a) A shape-anisotropic single-domain magnetostrictive nanomagnet and axis assignment. (b) The time-cycle of uniaxial stress (along $z$-direction) and magnetic field, \textit{H} (along $\theta=0^\circ$ direction) applied on the nanomagnet. (c) The in-plane ($y$-$z$ plane, $\phi=\pm90^\circ$) potential landscapes of the nanomagnet at different instants of time. (d) The complete three-dimensional potential landscapes of the nanomagnet at different instants of time. The solid lines on the landscapes correspond to the in-plane potential profiles, which are shown in (c). Consideration of these out-of-plane potential landscapes plays an important role in shaping the magnetization dynamics discussed later on. At time $t_1$, magnetization can reside in either of the wells ($\theta=0^\circ$ or $\theta=180^\circ$) with 50\% probability each, however, at time $t_2$, it reaches at $\theta=90^\circ$, i.e., the bit is erased. At time $t_3$ and onwards, magnetization reaches at $\theta \simeq 0^\circ$ deterministically.}
\label{fig:magnetization_potential}
\end{figure*}

\section{Model} We consider that the single-domain magnetostrictive nanomagnet is shaped like an elliptical cylinder  with its elliptical cross-section lying on the $y$-$z$ plane; the major axis and minor axis are aligned along the $z$- and $y$-direction, respectively (see Fig.~\ref{fig:magnetization_potential}a). In standard spherical coordinate system, $\theta$ is the polar angle and $\phi$ is the azimuthal angle. Any deviation of magnetization from magnet's plane ($y$-$z$ plane, $\phi = \pm90^{\circ}$) is termed as out-of-plane excursion. The dimensions of the major axis, the minor axis, and the thickness are $a$, $b$, and $l$, respectively. So the magnet's volume is $\Omega=(\pi/4)abl$. Due to this shape anisotropy, the two degenerate states $\theta=0^\circ$ and $\theta=180^\circ$ along the $z$-direction (easy axis) can store a bit of information. The $y$-axis is the in-plane hard axis and the $x$-axis is the out-of-plane hard axis. Since $l \ll b$, the out-of-plane hard axis is much harder than the in-plane hard axis.

Figure~\ref{fig:magnetization_potential}b shows the time-cycle of uniaxial stress (along $z$-direction) and a magnetic field $H$ (along $\theta=0^\circ$ direction) applied on the magnetostrictive nanomagnet. The adiabatic pulses should be slow enough that magnetization follows its potential landscape and the quasistatic assumption is valid. We can write the total energy of the nanomagnet per unit volume as the sum of three energies -- the shape anisotropy energy, anisotropies induced due to applied stress and asymmetry-making magnetic field (note that it is possible to replace the magnetic field by intrinsic interface coupling between polarization and magnetization in multiferroic heterostructures~\cite{RefWorks:649,roy14_2}) -- as follows~\cite{RefWorks:157}:

\begin{equation}
E_{total}(\theta,\phi,t) = B_{shape}(\phi) \, sin^2\theta  - B_{stress}(t)\,cos^2\theta \nonumber  - B_{asymm}(t) \, cos\,\theta,
\label{eq:E_total}
\end{equation}

where

\begin{eqnarray}
B_{shape}(\phi) &=& (1/2) \, M \, [H_k  + H_d\, cos^2\phi],\\
B_{stress}(t) &=& (3/2) \, \lambda_s \sigma(t),\\
B_{asymm}(t) &=& M H(t),
\label{eq:anisotropies}
\end{eqnarray}
\noindent
$M=\mu_0 M_s$, $\mu_0$ is permeability of free space, $M_s$ is the saturation magnetization, $H_k=(N_{dy}-N_{dz})\,M_s$ is the Stoner-Wohlfarth switching field~\cite{RefWorks:557}, $H_d=(N_{dx}-N_{dy})\,M_s$ is the demagnetization field~\cite{RefWorks:157}, $N_{dm}$ is the m$^{th}$ ($m=x,y,z$) component of the demagnetization factor~\cite{RefWorks:402} ($N_{dx} \gg N_{dy} > N_{dz}$), $(3/2) \lambda_s$ is the magnetostriction coefficient of the magnetostrictive nanomagnet~\cite{RefWorks:157}, $\sigma=Y\epsilon$ is the stress, $Y$ is the Young's modulus, $\epsilon$ is the strain that generates the stress, and $H(t)$ is the magnetic field at time $t$.

Note that a negative $\lambda_s \sigma$ product will favor alignment of the magnetization along the minor axis ($y$-axis) and according to our convention, in a material having \emph{positive} $\lambda_s$, a sufficiently high \emph{compressive} stress that can overcome the in-plane shape-anisotropic barrier will rotate the magnetization toward the in-plane hard axis ($\theta=90^\circ$, $\phi=\pm90^\circ$). This is depicted in Figs.~\ref{fig:magnetization_potential}c and~~\ref{fig:magnetization_potential}d that magnetization initially starting from any of the two degenerate stable states ($\theta=0^\circ$, state $0$ and $\theta=180^\circ$, state $1$ with probabilities, $p_0 = p_1 = 0.5$) comes to the in-plane hard axis at time $t_2$. The initial entropy of the system at time $t_1$ is $-k\sum_n p_n\,ln\,p_n = k\, ln(2)$. However, when stress is applied between time $t_1$ and $t_2$, the barrier separating the two stable states gets removed and the potential landscape becomes monostable in $\theta$-space (at $\theta=90^\circ$) with entropy zero. A reduction of entropy $k\, ln(2)$ must be dissipated as heat during this process. The reason that the barrier is removed before applying any asymmetry-making field $H$ is to make the erasure process independent of barrier height, which determines the hold failure probability and also to resist thermal fluctuations by making the monostable well deep enough.

Note that upon reaching $\theta=90^\circ$, magnetization may situate in $\phi$-space either at $\phi=+90^\circ$ or at $\phi=-90^\circ$, which is a consequence of having the planar shape of the nanomagnet, however, it should not be related to $\theta$-space in which the magnetization is switched. The $\phi$-space does exist even at $\theta \simeq 0^\circ$ or $\theta \simeq 180^\circ$ [e.g., see Supplementary Fig. S1, where there exist two peaks at $\phi=\pm 90^\circ$ around $\theta \simeq 180^\circ$].

Between times $t_2$ and $t_3$, an asymmetry-making field $H$ is applied to deterministically rotate the magnetization to state $\theta=0^\circ$. The degree of this tilt should be sufficient enough to dissuade thermal fluctuations. Afterwards, both the stress and $H$ are removed to complete the switching process. Note that the entropy of the system at time $t_2$ and onwards are zero. Hence there is no minimum energy dissipation bound when magnetization traverses from time $t_2$ to $t_3$ as the dissipation can be made arbitrarily small.

The torque acting on the magnetization is derived from the gradient of potential landscape. Additionally, there is a random thermal field [$\mathbf{h}(t) =\sum_i h_i(t) \, \mathbf{\hat{e}_i}$, $i \in (x,y,z)$] to incorporate thermal fluctuations~\cite{RefWorks:186}. Solving the stochastic Landau-Lifshitz-Gilbert equation~\cite{RefWorks:162,RefWorks:161,RefWorks:186} of magnetization dynamics analytically (see Supplementary material for details), we get the following coupled equations for $\theta$ and $\phi$:

\begin{eqnarray}
\left(1+\alpha^2 \right) \frac{d\theta}{dt} &=& \frac{|\gamma|}{M} \lbrack  - \alpha B(\phi,t) sin(2\theta) - \alpha B_{asymm}(t)\, sin\theta \nonumber \\
				&& + B_{shape,\phi}(\phi) sin\theta + \left(\alpha P_\theta(\theta,\phi,t) + P_\phi (\theta,\phi,t) \right) \rbrack,
 \label{eq:theta_dynamics_thermal}
\end{eqnarray}
\begin{eqnarray}
\left(1+\alpha^2 \right) \frac{d \phi}{dt} &=& \frac{|\gamma|}{M} \lbrack 2 B(\phi,t) cos\theta + B_{asymm}(t) \nonumber\\
					&& + \alpha B_{shape,\phi}(\phi) - {[sin\theta]^{-1}} \left(P_\theta (\theta,\phi,t) - \alpha P_\phi (\theta,\phi,t) \right) \rbrack \nonumber \\
	&& \hspace*{5.5cm} (sin\theta \neq 0),
\label{eq:phi_dynamics_thermal}
\end{eqnarray}

where

\begin{eqnarray}
B(\phi,t) &=& B_{shape}(\phi) + B_{stress}(t), \\
B_{shape,\phi}(\phi) &=& (1/2) \, M H_d \,  sin(2\phi), \label{eq:B_shape_phi}\\
P_\theta(\theta,\phi,t) &=& M  \lbrack h_x(t)\,cos\theta\,cos\phi + h_y(t)\,cos\theta sin\phi - h_z(t)\,sin\theta \rbrack, \\
P_\phi(\theta,\phi,t) &=& M  \lbrack h_y(t)\,cos\phi -h_x(t)\,sin\phi\rbrack, \\
h_i(t) &=& \sqrt{\frac{2 \alpha kT}{|\gamma| M \Omega \Delta t}} \, G_{(0,1)}(t) \quad (i=x,y,z), 
\label{eq:LLG_parts}
\end{eqnarray}
\noindent
$\alpha$ is the dimensionless phenomenological Gilbert damping parameter~\cite{RefWorks:651,RefWorks:647,RefWorks:650,RefWorks:425}, $\gamma$ is the gyromagnetic ratio for electrons, $1/\Delta t$ is proportional to the attempt frequency of the thermal field, $\Delta t_{sim} = 10^3 \, \Delta t$ is the simulation time-step used, and the quantity $G_{(0,1)}(t)$ is a Gaussian distribution with zero mean and unit variance.

The energy dissipated in the nanomagnet due to Gilbert damping can be expressed as  $E_d = \int_0^{\tau}P_d(t) dt$, where $\tau$ is the time taken during the erasure cycle [i.e., $t_5-t_1$ in Fig.~\ref{fig:magnetization_potential}(b)], and $P_d(t)$ is the power dissipated at time $t$ per unit volume given by

\begin{equation}
P_d(t) = \frac{\alpha \, |\gamma|}{(1+\alpha^2) M} \, |\mathbf{T_E} (\theta(t), \phi(t), t)|^2,
\label{eq:power_dissipation}
\end{equation}
\noindent
where $\mathbf{T_E}$ is the torque (see Supplementary material for details) acting on the magnetization due to the gradient of potential energy $E_{total}$ [Equation~\eref{eq:E_total}]. Thermal field with mean zero does not cause any net energy dissipation but it causes variability in the energy dissipation by scuttling the trajectory of magnetization.

Assuming magnetization resides in-plane ($\phi=\pm90^\circ$) between times $t_1$ and $t_2$, $(3/2)\lambda_s \sigma_{min} = - (1/2) M H_k$, $\sigma_{min}$ is the stress at which the energy barrier is just removed, and $\sigma_{max} = \sigma(t_2) = 2 \sigma_{min}$, we can deduce from Equation~\eref{eq:theta_dynamics_thermal} without considering thermal fluctuations (i.e., $\mathbf{h}=0$) that

\begin{eqnarray}
T = t_2 - t_1 &=& \frac{8\,(1+\alpha^2)}{\alpha |\gamma| H_k} \int_{\theta_1=\theta(t_1)}^{\theta_2=\theta(t_2)} \frac{d\theta}{sin(2\theta)} \nonumber\\
							&=& \frac{4\,(1+\alpha^2)}{\alpha |\gamma| H_k} \, ln \left( \frac{tan\theta_2}{tan\theta_1} \right).
\label{eq:stress_2D}
\end{eqnarray}

Note that at $\theta_1=180^\circ$ and $\theta_2=90^\circ$, the above expression is undefined or it would take exceedingly large amount of time to start \emph{exactly} from $\theta=180^\circ$ and to reach \emph{exactly} at $\theta=90^\circ$ since the torque acting on the magnetization at these points due the gradient of potential landscape vanishes. Fortunately, thermal fluctuations can prevent this lockjam in such circumstances. Assuming $\theta_1=175^\circ$ and $\theta_2=95^\circ$, $ln(tan\theta_2/tan\theta_1)=4.8725$. This closed-form expression gives us an estimate over the time required for magnetization to reach $\theta\simeq 90^\circ$ depending on the material parameters involved.

It needs mention here that particularly in the presence of thermal fluctuations, magnetization may temporarily traverse on higher potential (not only in-plane of the nanomagnet but also out-of-plane, i.e., when $\phi \neq \pm90^\circ$) and dissipate energy when it comes back to lower potential. A less than one degree of  deflection in the out-of-plane direction can have immense consequence on magnetization dynamics. The key reason behind is that the out-of-plane demagnetization field $H_d$ is about a couple of orders of magnitude higher than $H_k$. When the magnetization deflects out-of-plane due to torque exerted on it, an additional torque of comparatively very high magnitudes comes into play [see Equation~\eref{eq:B_shape_phi}], which makes the dynamics \emph{fast}. It is true that the out-of-plane excursion causes power dissipation but switching also becomes fast, so that the net energy dissipation may be smaller compared to the case when it is \emph{assumed} that magnetization is confined to the magnet's plane. Since the Landauer limit of energy dissipation is very small, we should particularly take into consideration this significant effect due to out-of-plane excursion of magnetization.

We will now analyze the magnetization dynamics between times $t_2$ and $t_3$. Assuming  $(3/2)\lambda_s \sigma_{max} = - M H_k$ and $\phi=\pm90^\circ$, the total energy becomes $E_{total}=-(1/2)M H_k\,sin^2\theta-M H(t)\,cos\theta$. We see that as $H$ goes from $0$ to $H_k$, the minimum value of $\theta$ goes from $\theta=90^\circ$ to $\theta=0^\circ$ continuously following the expression $\theta_{min}(t)=cos^{-1}(H(t)/H_k)$. 

Due to the smooth transitions of magnetization, with adiabatic pulses between times $t_1$ and $t_3$, there is no lower bound of energy dissipation \emph{in the absence of thermal fluctuations}. However, the Landauer principle remains intact since this is $T=0\,K$ case and thus the energy dissipation proportional to temperature is also zero. If we incorporate random thermal fluctuations at finite temperature, magnetization will get deflected uphill in the potential landscape even with very slow ramp of pulses and will incur energy dissipation, which is subjected to Landauer bound.

\section{Results} We consider the magnetostrictive nanomagnet to be made of polycrystalline Galfenol (FeGa), which has the following material properties -- magnetostrictive coefficient ($(3/2)\lambda_s$): +150$\times$10$^{-6}$, saturation magnetization ($M_s$):  8$\times$10$^5$ A/m, Gilbert damping parameter ($\alpha$): 0.025, and  Young's modulus (Y): 140 GPa~\cite{RefWorks:167,RefWorks:801}. The dimensions of the nanomagnet is chosen as $a$ = 100 nm, $b$ = 90 nm, and $l$ = 6 nm, which ensures that the nanomagnet has a single ferromagnetic domain~\cite{RefWorks:133}. With the chosen dimensions, the Stoner-Wohlfarth switching field $H_k$ becomes $\sim$$0.01\,M_s$. The values of stress ($\sigma_{max}$), strain ($\epsilon$), and asymmetric field ($H$) are 60.6 MPa, 433$\times$10$^{-6}$, and 0.01 T, respectively. 


\begin{figure*}[htbp]
\centering
\includegraphics[width=170mm]{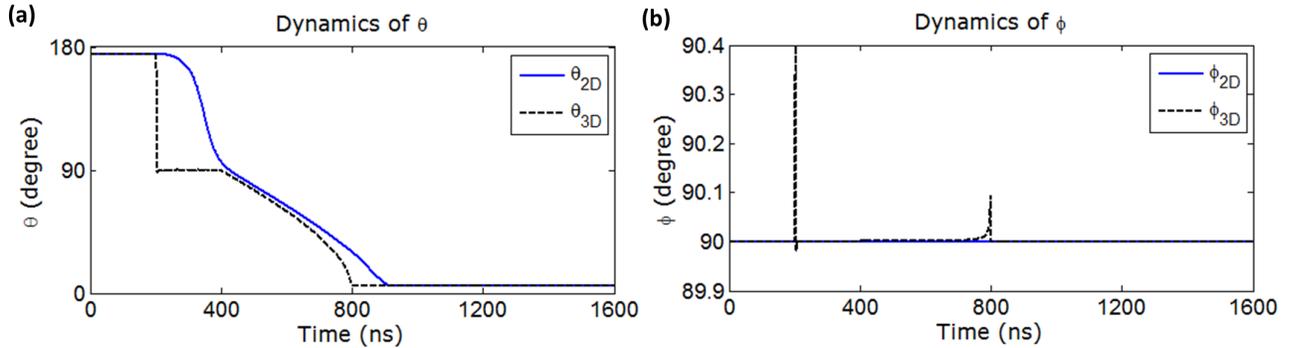}
\caption{Magnetization dynamics when $t_{n+1} - t_{n}$ = 400 ns ($n$=1-4). Thermal fluctuations are not incorporated during this simulation, however, since exactly at $\theta=180^\circ$, the torque acting on the magnetization is zero, an initial 5 degree of deflection for $\theta$ is chosen. The initial value of $\phi$ is $90^\circ$, choice of the other in-plane angle $\phi=270^\circ$ ($-90^\circ$) is analogous. Both the cases when magnetization is \emph{assumed} to be confined on magnet's plane (2D) and that when considering the full three-dimensional potential landscape (3D) are plotted. (a) $\theta$-dynamics, and (b) $\phi$-dynamics.}
\label{fig:dynamics_0K_theta_phi}
\end{figure*}

Figure~\ref{fig:dynamics_0K_theta_phi} shows the magnetization dynamics when both the ramp-up and ramp-down times of stress and \emph{H} field are 400 ns (Equation~\eref{eq:stress_2D} gives a value of $\sim$400 ns). Both the results -- when magnetization is \emph{assumed} to be restricted on the magnet's plane (2D \emph{assumption}) and while considering full there-dimensional potential landscape -- are plotted. Note that even a less than one degree deflection of magnetization out of magnet's plane (see azimuthal angle $\phi$-plots in Fig.~\ref{fig:dynamics_0K_theta_phi}b) has caused a significant change in the $\theta$-dynamics due to inherent $\theta$-$\phi$ coupling in LLG equation. The \emph{actual} dynamics is quite faster than when 2D \emph{assumption} is enforced. In small systems, the out-of-plane demagnetization factor $N_{dx}$ is not that high compared to the $y$- and $z$-component (ratio is about 10 times) and since there are torques acting on the magnetization in the $\phi$-direction, magnetization can easily deflect out-of-plane particularly in the presence of thermal fluctuations. 

\begin{figure}[htbp]
\centering
\includegraphics[width=3.4in]{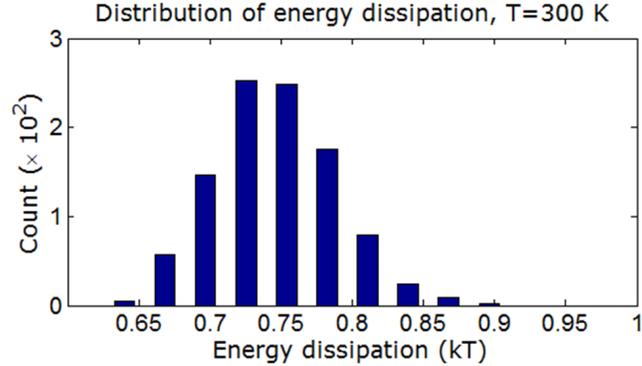}
\caption{Distribution of energy dissipation at room-temperature (300 K) during the time interval $t_2 - t_1$ when stress is brought up from zero to a maximum value decreasing the entropy of the system by $k\,ln(2)=0.6932\,k$. A concomitant amount of energy must be dissipated according to Landauer's principle. Note that for a few cases, the Landauer bound is violated but the mean of the energy dissipation ($0.74\,kT$) exceeds the Landauer bound safeguarding the Landauer's principle and the second law of thermodynamics. A moderately large (1000) number of simulations in the presence thermal fluctuations have been performed to generate this distribution.}
\label{fig:distribution_energy_300K}
\end{figure}

Figure~\ref{fig:distribution_energy_300K} shows the distribution of energy dissipation at 300 K during the time interval $t_2 - t_1$ = 100 ns to depict that the \emph{generalized} Landauer principle of energy dissipation is maintained. Stochastic violation of Landauer's bound 0.6932$\,kT$ due to thermal fluctuations is observed but the \emph{mean} energy dissipation does respect the Landauer's bound. In the presence of thermal fluctuations, we note that a slower ramp rate of pulse is sufficient than that for 0 K, since thermal fluctuations does a \emph{net} favor to direct magnetization going downhill. The mean energy dissipation during the time interval $t_3 - t_2$ (when magnetization traverses from $\theta=90^\circ$ towards $\theta=0^\circ$) is 0.07$\,kT$, which does not have any bound since there is no entropy loss in the system. 


\section{Conclusions} We have shown that Landauer limit of energy dissipation is achievable in a magnetostrictive nanomagnet. With realistic parameters and an well-accepted model, i.e., Landau-Lifshitz-Gilbert equation of magnetization dynamics, we have performed simulations and the results show that unlike the case of a particle with linear motion, the out-of-plane excursion of magnetization plays a crucial role in shaping the magnetization dynamics while achieving the Landauer limit. We hope that our finding would stimulate experimental efforts to demonstrate that the magnetostrictive nanomagnets are suitable for exploring the thermodynamic limit of energy dissipation. On applied perspective, such miniscule dissipation can be the basis of ultra-low-energy computing for our future information processing systems. Also, this can open up unprecedented applications that need to work with the energy harvested from the environment e.g., monitoring an epileptic patient's brain to report an impending seizure by drawing energy solely from the patient's body.

\section*{References}

\providecommand{\newblock}{}

\end{document}


\title[]{Supplementary Information\\Landauer limit of energy dissipation in a magnetostrictive particle}

\author{Kuntal Roy}
\address{School of Electrical and Computer Engineering, Purdue University, West Lafayette, Indiana 47907, USA\\
	* Some works for this paper were performed prior to joining Purdue University.}

\ead{royk@purdue.edu}

\date{\today}

\section{Solution of the stochastic Landau-Lifshitz-Gilbert (LLG) equation} We consider that the single-domain magnetostrictive nanomagnet is shaped like an elliptical cylinder  with its elliptical cross-section lying on the $y$-$z$ plane; the major axis and minor axis are aligned along the $z$- and $y$-direction, respectively (see Fig.~1a in the main Letter). In standard spherical coordinate system, $\theta$ is the polar angle and $\phi$ is the azimuthal angle. Any deviation of magnetization from magnet's plane ($y$-$z$ plane, $\phi = \pm90^{\circ}$) is termed as out-of-plane excursion. The dimensions of the major axis, the minor axis, and the thickness are $a$, $b$, and $l$, respectively. So the magnet's volume is $\Omega=(\pi/4)abl$. Due to this shape anisotropy, the two degenerate states $\theta=0^\circ$ and $\theta=180^\circ$ along the $z$-direction (easy axis) can store a bit of information. The $y$-axis is the in-plane hard axis and the $x$-axis is the out-of-plane hard axis. Since $l \ll b$, the out-of-plane hard axis is much harder than the in-plane hard axis.

We can write the total energy of the nanomagnet per unit volume as the sum of three energies -- the shape anisotropy energy, anisotropies induced due to applied stress and magnetic field (note that we assume a \emph{polycrystalline} nanomagnet, so that we ignore the magnetocrystalline anisotropy energy) -- as follows:

\begin{eqnarray}
E_{total}(\theta,\phi,t) &=& E_{shape}(\theta,\phi) + E_{stress}(\theta,t) + E_{asymm}(\theta,t) \nonumber\\
												 &=& B_{shape}(\phi) \, sin^2\theta - B_{stress}(t)\,cos^2\theta - B_{asymm}(t) \, cos\,\theta,
\label{eq:total_anisotropy}
\end{eqnarray}
where
\begin{eqnarray}
B_{shape}(\phi) &=& (1/2) \, M \, [H_k  + H_d\, cos^2\phi],\\
B_{stress}(t) &=& (3/2) \, \lambda_s \sigma(t),\\
B_{asymm}(t) &=& M H(t),
\label{eq:anisotropies}
\end{eqnarray}
\noindent
$M=\mu_0 M_s$, $\mu_0$ is permeability of free sapce, $M_s$ is the saturation magnetization, $H_k=(N_{dy}-N_{dz})\,M_s$ is the Stoner-Wohlfarth switching field~\cite{RefWorks:157}, $H_d=(N_{dx}-N_{dy})\,M_s$ is the demagnetization field~\cite{RefWorks:157}, $N_{dm}$ is the m$^{th}$ ($m=x,y,z$) component of the demagnetization factor~\cite{RefWorks:402} ($N_{dx} \gg N_{dy} > N_{dz}$), $(3/2) \lambda_s$ is the magnetostriction coefficient of the magnetostrictive nanomagnet~\cite{RefWorks:157}, $\sigma(t)$ is the stress, and $H(t)$ is the magnetic field, at time $t$.

Note that a negative $\lambda_s \sigma(t)$ product will favor alignment of the magnetization along the minor axis ($y$-axis) and according to our convention, in a material having \emph{positive} $\lambda_s$, a sufficiently high \emph{compressive} stress that can overcome the in-plane shape-anisotropic barrier will rotate the magnetization toward the in-plane hard axis ($\theta=90^\circ$, $\phi=\pm90^\circ$).

The magnetization \textbf{M} of the nanomagnet has a constant magnitude but a variable direction, so that we can represent it by a vector of unit norm $\mathbf{n_m} =\mathbf{M}/|\mathbf{M}| = \mathbf{\hat{e}_r}$ where $\mathbf{\hat{e}_r}$ is the unit vector in the radial direction in spherical coordinate system represented by ($r$,$\theta$,$\phi$). The other two unit vectors in the spherical coordinate system are denoted by $\mathbf{\hat{e}_\theta}$ and $\mathbf{\hat{e}_\phi}$ for $\theta$ and $\phi$ rotations, respectively. 

The effective field and torque acting on the magnetization per unit volume due to gradient of potential landscape can be expressed as $\mathbf{H_{eff}} = - \nabla E_{total} = - (\partial E_{total}/\partial \theta)\,\mathbf{\hat{e}_\theta} - (1/sin\theta)\,(\partial E_{total}/\partial \phi)\,\mathbf{\hat{e}_\phi}$ and $\mathbf{T_E}= \mathbf{n_m} \times \mathbf{H_{eff}}$, respectively.
\begin{eqnarray}
\mathbf{T_E} (\theta,\phi,t) &&= - [B(\phi,t)\, sin(2\theta) + B_{asymm}(t)\,sin\theta] \,\mathbf{\hat{e}_\phi} - B_{shape,\phi}(\phi)\,sin\theta \,\mathbf{\hat{e}_\theta}, \nonumber\\
&&
\label{eq:torque}
\end{eqnarray}
\noindent
where 
\begin{eqnarray}
B(\phi,t) &=& B_{shape}(\phi) + B_{stress}(t), \\
B_{shape,\phi}(\phi) &=& (1/2) \, M H_d \,  sin(2\phi). 
\label{eq:B_Bshape_phi}
\end{eqnarray}
 
The effect of random thermal fluctuations is incorporated via a random magnetic field $\mathbf{h}(t)= h_x(t)\mathbf{\hat{e}_x} + h_y(t)\mathbf{\hat{e}_y} + h_z(t)\mathbf{\hat{e}_z}$, where $h_i(t)$ ($i=x,y,z$) are the three components of the random thermal field in Cartesian coordinates. We assume the properties of the random field $\mathbf{h}(t)$ as described in Ref.~\cite{RefWorks:186}. The random thermal field can be written as~\cite{RefWorks:186}
\begin{equation}
h_i(t) = \sqrt{\frac{2 \alpha kT}{|\gamma| M \Omega \Delta t}} \; G_{(0,1)}(t) \quad (i \in x,y,z)
\label{eq:ht}
\end{equation}
\noindent
where $\alpha$ is the dimensionless phenomenological Gilbert damping parameter, $\gamma$ is the gyromagnetic ratio for electrons, $1/\Delta t$ is proportional to the attempt frequency of the thermal field, $\Delta t_{sim} = 10^3 \, \Delta t$ is the simulation time-step used, and the quantity $G_{(0,1)}(t)$ is a Gaussian distribution with zero mean and unit variance. 

The thermal field and corresponding torque acting on the magnetization per unit volume can be written as $\mathbf{H_{TH}}(\theta,\phi,t)=P_\theta(\theta,\phi,t)\,\mathbf{\hat{e}_\theta}+P_\phi(\theta,\phi,t)\,\mathbf{\hat{e}_\phi}$ and $\mathbf{T_{TH}}(\theta,\phi,t)=\mathbf{n_m} \times \mathbf{H_{TH}}(\theta,\phi,t)$, respectively, 
where
\begin{eqnarray}
P_\theta(\theta,\phi,t) &=& M  \lbrack h_x(t)\,cos\theta\,cos\phi + h_y(t)\,cos\theta sin\phi - h_z(t)\,sin\theta \rbrack,\\
P_\phi(\theta,\phi,t) &=& M  \lbrack h_y(t)\,cos\phi -h_x(t)\,sin\phi\rbrack.
\label{eq:thermal_parts}
\end{eqnarray}
\noindent

The magnetization dynamics under the action of the torques $\mathbf{T_{E}}(t)$ and 
$\mathbf{T_{TH}}(t)$ is described by the stochastic Landau-Lifshitz-Gilbert (LLG) equation~\cite{RefWorks:162,RefWorks:161,RefWorks:186} as follows.
\begin{equation}
\frac{d\mathbf{n_m}}{dt} - \alpha \left(\mathbf{n_m} \times \frac{d\mathbf{n_m}}{dt} \right) = -\frac{|\gamma|}{M} \left\lbrack \mathbf{T_E} +  \mathbf{T_{TH}}\right\rbrack.
\end{equation}

Solving the above equation analytically, we get the following coupled equations of magnetization dynamics for $\theta$ and $\phi$:
\begin{eqnarray}
\left(1+\alpha^2 \right) \frac{d\theta}{dt} &=& \frac{|\gamma|}{M} \lbrack  - \alpha B(\phi,t) sin(2\theta) - \alpha B_{asymm}(t)\, sin\theta \nonumber \\
				&& + B_{shape,\phi}(\phi) sin\theta + \left(\alpha P_\theta(\theta,\phi,t) + P_\phi (\theta,\phi,t) \right) \rbrack,
 \label{eq:theta_dynamics}
\end{eqnarray}
\begin{eqnarray}
\left(1+\alpha^2 \right) \frac{d \phi}{dt} &=& \frac{|\gamma|}{M} \lbrack 2 B(\phi,t) cos\theta + B_{asymm}(t) \nonumber\\
					&& + \alpha B_{shape,\phi}(\phi) - {[sin\theta]^{-1}} \left(P_\theta (\theta,\phi,t) - \alpha P_\phi (\theta,\phi,t) \right) \rbrack \nonumber \\
	&& \hspace*{5.5cm} (sin\theta \neq 0),
\label{eq:phi_dynamics}
\end{eqnarray}
We need to solve the above two coupled equations numerically to track the trajectory of magnetization over time, in the presence of thermal fluctuations.

The energy dissipated in the nanomagnet due to Gilbert damping can be expressed as  $E_d = \int_0^{\tau}P_d(t) dt$, where $\tau$ is the switching delay and $P_d(t)$ is the power dissipated at time $t$ per unit volume given by
\begin{equation}
P_d(t) = \frac{\alpha \, |\gamma|}{(1+\alpha^2) M} \, |\mathbf{T_E} (\theta(t), \phi(t), t)|^2 .
\label{eq:power_dissipation}
\end{equation}
Thermal field with mean zero does not cause any net energy dissipation but it causes variability in the energy dissipation by scuttling the trajectory of magnetization.

\vspace*{5mm}

\section{Distribution of initial orientation of magnetization} When the magnetization direction is {\it exactly} along the easy axis, i.e., $\sin\theta=0$ ($\theta=0^\circ$ or $\theta=180^\circ$), the torque acting on the magnetization given by Equation~\eref{eq:torque} becomes zero. That is why only thermal fluctuations can deflect the magnetization vector \emph{exactly} from the easy axis. Consider the situation when $\theta=180^\circ$. From Equations~\eref{eq:theta_dynamics} and~\eref{eq:phi_dynamics}, we get~\cite{roy13_2}
\begin{equation}
\phi(t) = tan^{-1} \left( \frac{\alpha h_y(t) + h_x(t)}{h_y(t) - \alpha h_x(t)} \right),
\label{eq:phi_dynamics_thermal}
\end{equation}
\begin{equation}
\frac{d\theta(t)}{dt} =  \frac{-|\gamma| (h_x^2(t) + h_y^2(t))}{\sqrt{(h_y(t)-\alpha h_x(t))^2 + (\alpha h_y(t) + h_x(t))^2}}.
\label{eq:theta_dynamics_thermal}
\end{equation}
\noindent
From the Equation~\eref{eq:theta_dynamics_thermal}, we notice clearly that the thermal fluctuations can deflect the magnetization \emph{exactly} from the easy axis since the time rate of change of $\theta(t)$ is non-zero in the presence of thermal fluctuations. Note that $d\theta(t)/dt$ does not depend on the component of the random thermal field along the $z$-axis, i.e. $h_z(t)$, which is a consequence of having $z$-axis as the easy axis of the nanomagnet. However, once the magnetization direction is even slightly deflected from the easy axis, all three components of the random thermal field along the $x$-, $y$-, and $z$-direction would come into play.

\begin{figure*}[htbp]
\centering
\subfigure[]{\includegraphics[width=3.4in]{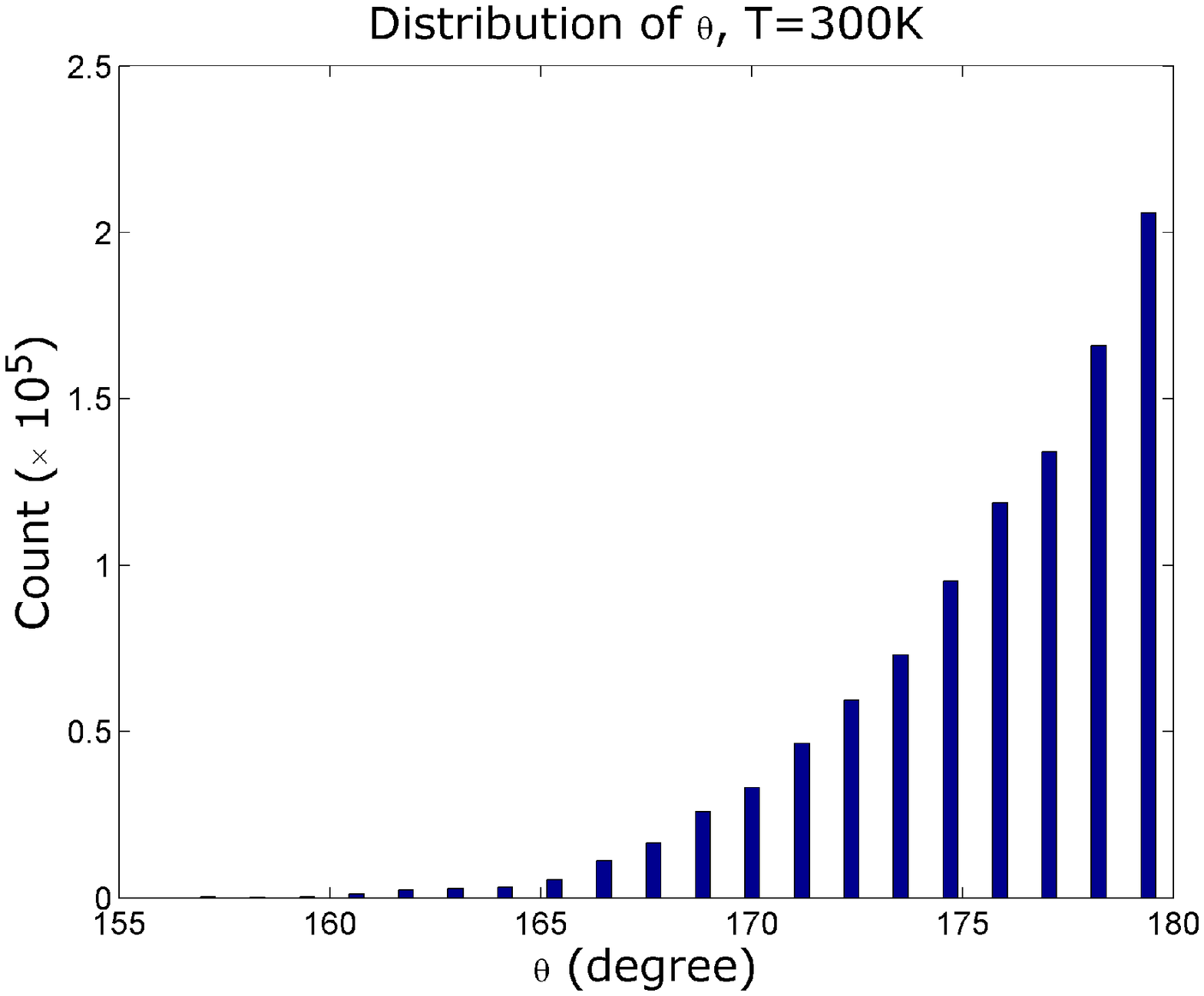}%
\label{fig:theta_distribution_T_300K}}
\hfil
\subfigure[]{\includegraphics[width=3.4in]{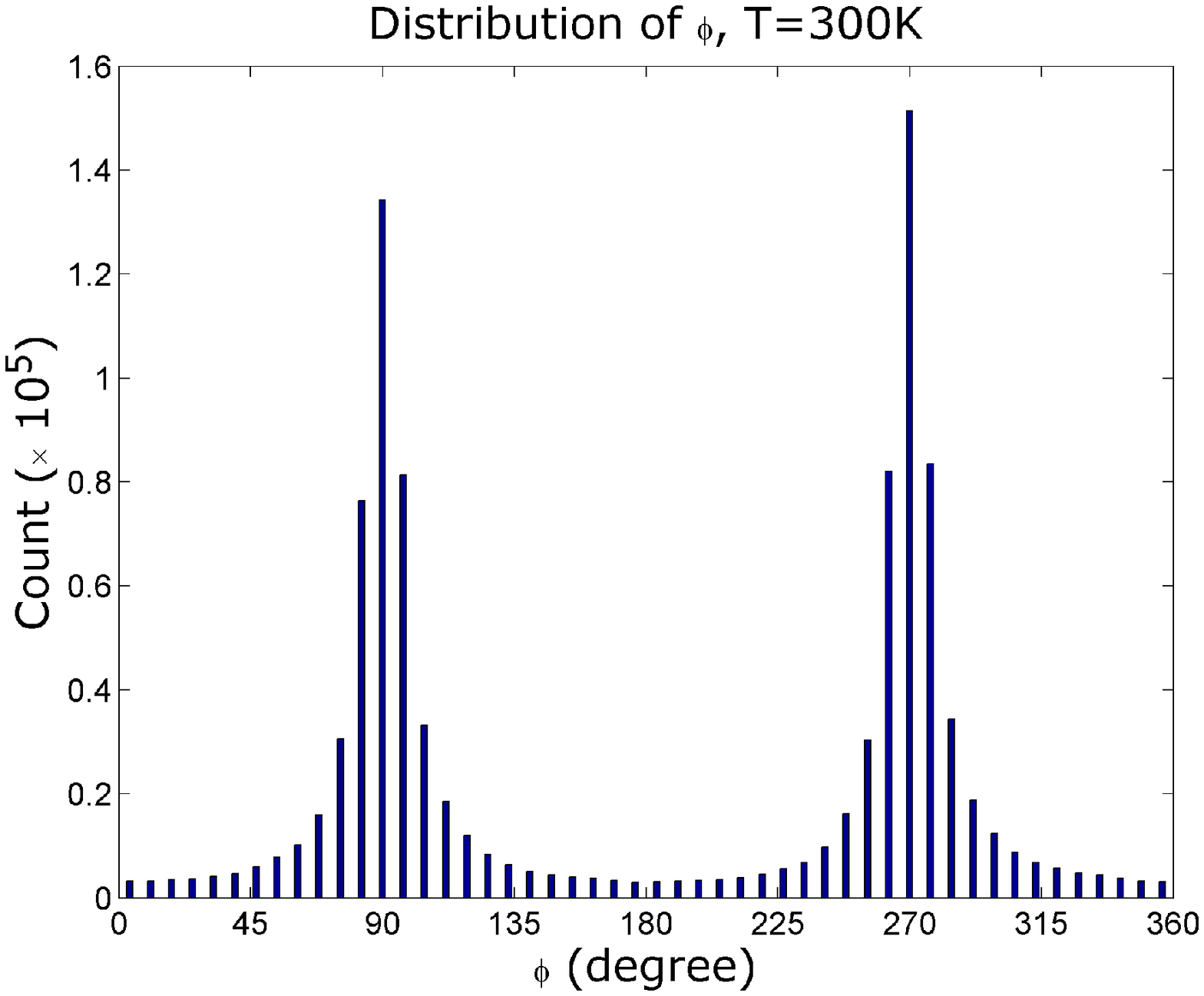}%
\label{fig:phi_distribution_T_300K}}
\caption{\label{fig:distribution_theta_phi_T_300K}Distribution of polar angle $\theta$ and azimuthal angle $\phi$ due to room-temperature (300 K) thermal fluctuations~\cite{roy13_2}.
(a) Distributions of the polar angle $\theta$. This is a Boltzmann distribution with mean value $\sim$175$^\circ$, while the most likely value is 180$^{\circ}$. 
(b) Distribution of the azimuthal angle $\phi$. These are two Gaussian distributions with peaks centered at $90^\circ$ and $270^\circ$ 
(or $-90^{\circ}$), which depicts that the most likely location of the magnetization vector is in the plane of the nanomagnet.}
\end{figure*}

When neither stress nor asymmetry-making $H$ field is applied on the magnetostrictive nanomagnet, magnetization just fluctuates around an easy axis provided that the shape anisotropy energy barrier is enough high to prevent spontaneous reversal of magnetization from one stable state to another in a short period of time. We have solved the Equations~\eref{eq:theta_dynamics} and~\eref{eq:phi_dynamics} while setting $\sigma = H = 0$ to track the dynamics of magnetization due to thermal fluctuations. So this will yield the distribution of the magnetization vector's initial orientation when stress or $H$ is turned on. Since the most probably value of magnetization is along an easy axis, the $\theta$-distribution is Boltzmann peaked at $\theta$ = 0$^{\circ}$ or 180$^{\circ}$. The $\phi$-distribution is Gaussian peaked at $\phi = \pm 90^{\circ}$ since these positions are minimum energy positions. Taking the initial distribution into account does introduce variability in the metrics calculated, however, it does not change the mean value significantly since the odds are \emph{exponentially} suppressed.



\section*{References}

\providecommand{\newblock}{}